\documentclass[runningheads]{llncs}
\usepackage{float}
\usepackage{caption}
\usepackage{subcaption}
\usepackage{graphicx}
\graphicspath{ {./pic/} }
\usepackage{listings}
\usepackage{comment}

\usepackage{amssymb,amsmath}
\usepackage{xcolor}

\newcommand{\mynote}[3]{
    \fbox{\bfseries\sffamily\scriptsize#1}
    {\small$\blacktriangleright$\textsf{\emph{\color{#3}{#2}}}$\blacktriangleleft$}}

\newcommand{\ik}[1]{\mynote{Ilya}{#1}{magenta}}

\begin{document}
\title{Execution of NVRAM Programs with Persistent Stack}
%
%
\author{Vitaly Aksenov\inst{1} \and
Ohad Ben-Baruch\inst{2} \and
Danny Hendler\inst{3} \and
Ilya Kokorin\inst{4} \and
Matan Rusanovsky\inst{5}}
\authorrunning{V. Aksenov et al.}
%
\institute{
ITMO University, \emph{Addr:} 49, Kronverksky ave, Saint-Petersburg, Russia, 197101; \emph{Email:} aksenov.vitaly@gmail.com; \emph{Phone:} +79516623399
\and
Ben-Gurion University, \emph{Addr}: P.O.B. 653 Beer-Sheva, Israel, 8410501; \emph{Email:} bbohad@gmail.com; \emph{Phone:} +97286461600
\and
Ben-Gurion University, \emph{Addr}: P.O.B. 653 Beer-Sheva, Israel, 8410501; \emph{Email:} hendlerd@bgu.ac.il; \emph{Phone:} +97286461600
\and
ITMO University, \emph{Addr:} 49, Kronverksky ave, Saint-Petersburg, Russia, 197101; \emph{Email:} kokorin.ilya.1998@gmail.com; \emph{Phone:} +79811639149
\and
Ben-Gurion University, \emph{Addr}: P.O.B. 653 Beer-Sheva, Israel, 8410501; \emph{Email:} matanru@post.bgu.ac.il; \emph{Phone:} +97286461600
}
\maketitle              
\begin{abstract}
Non-Volatile Random Access Memory (NVRAM) is a novel type of hardware that combines the benefits of traditional persistent memory (persistency of data over hardware failures) and DRAM (fast random access). In this work, we describe an algorithm that can be used to execute NVRAM programs and recover the system after a hardware failure while taking the architecture of real-world NVRAM systems into account. Moreover, the algorithm can be used to execute NVRAM-destined programs on commodity persistent hardware, such as hard drives. That allows us to test NVRAM algorithms using only cheap hardware, without having access to the NVRAM. We report the usage of our algorithm to implement and test NVRAM CAS algorithm.

\end{abstract}

\section{Introduction}

For a long time the industry assumed the existence of two distinct types of the memory. The first one is a persistent memory that preserves its content even in the presence of hardware (e.g., power) failures. This type of memory was assumed to support mainly sequential block access with the poor performance of random access. Due to its ability to persist data this kind of memory is widely used to recover the system after a hardware failure: one can load the data from the persistent memory and restore the state of the application before the crash. The second type of the memory is DRAM that supports fast random byte-addressable access but loses its content on hardware failures. Due to its speed, this kind of memory is widely used in high-performance computations.

Nowadays, we can get benefits from both of these worlds due to the invention of Non-Volatile Random Access Memory (NVRAM)---a novel type of hardware that combines both the persistency and the fast random access. This allows us to implement low-latency persistent data structures that require random access to the memory, e.g., binary search trees, linked lists, and etc. A lot of work has been done to come up with data structures, hand-tuned for the NVRAM \cite{david2018log,chen2015persistent,friedman2018persistent}. Some authors propose techniques, that can be used to transform DRAM-resident data structures into the ones suitable for the NVRAM \cite{chauhan2016nvmove,friedman2020nvtraverse}.

Despite the speed of the NVRAM is compatible with the speed of the DRAM, the NVRAM is not expected to replace volatile memory totally since processor registers and the NVRAM cache are expected to remain volatile. Thus, even on NVRAM systems, a system failure leads to: 1)~the loss of the results of recent computations since \texttt{x86} computations are performed using volatile processor registers, and 2)~the loss of data that was written to the NVRAM cache and has not been flushed to the NVRAM.

To make sure that the written data becomes persistent, we should flush one or several cache lines to the NVRAM. Flush of a single cache line is an atomic action: if a crash occurs during cache line flushing, the whole cache line is either persisted or not. However, if we want to flush multiple cache lines at a time, a crash event can occur between flushes~--- in such a case, only a part of the data becomes persistent while the rest is lost.

This yields one of the major challenges of NVRAM. If a system failure happens during a complex update, when some updated values have been flushed to the NVRAM from the cache while others still reside in the cache, non-flushed memory is lost and after the restart the NVRAM appears to be in an inconsistent state. 

Due to the difficulty of ensuring storage consistency in the presence of the volatile NVRAM cache, a lot of works assume the absence of such cache \cite{attiya2018nesting,ben2019delay,berryhill2016robust,blelloch2018parallel}. However, in this work we consider real-life systems, thus we take the volatility of the NVRAM cache into account.

Another problem with the NVRAM is defining which executions are considered ``correct'' in the presence of hardware failures, that can lead to the loss of data. Despite a lot of correctness conditions were defined in the previous years \cite{aguilera2003strict,berryhill2016robust,guerraoui2004robust,izraelevitz2016linearizability,attiya2018nesting}, only \emph{Nesting Safe Recoverable Linearizability} \cite{attiya2018nesting} describes the work with nested functions. Thus, maintaining persistent call stack is a crucial part of systems based on this concept. However, while methods of maintaining NVRAM heap are well-studied \cite{cai2020understanding,bhandari2016makalu}, methods of maintaining the persistent program stack are not studied at all: other works just assume the existence of a persistent call stack \cite{attiya2018nesting,ben2019delay,friedman2020nvtraverse}. 

Moreover, the persistent stack allows us to design and implement novel complex system recovery algorithms, which can be faster than traditional log-based system recovery methods. Previously, such complex algorithms were considered impractical for traditional persistent memory systems due to the high latency of random access of traditional persistent memory, following directly from its mechanical nature, but on NVRAM-based systems such complex algorithms may be found useful.

In this work, we describe an algorithm, based on the implementation of the persistent call stack, that can be used to execute NVRAM programs and recover the system after a hardware failure while taking the architecture of real-world NVRAM systems into account. Moreover, the algorithm can be used to execute NVRAM-destined programs on commodity persistent hardware, such as hard drives. That allows us to test NVRAM algorithms using only cheap persistent hardware, such as HDD, SSD, etc., without having access to the NVRAM. We report the usage of our algorithm to implement and test correct and incorrect versions of the NVRAM CAS algorithm~\cite{attiya2018nesting}. Also, we describe a method, that can be used to verify executions of NVRAM CAS algorithm for serializability. 

The rest of the work is organized as follows. In Section~\ref{model-section}, we discuss the system model, various failure models, operation execution model and talk about different correctness conditions, suitable for the NVRAM. In Section~\ref{stack-section}, we discuss the concept of the persistent program stack and its implementation. In section~\ref{system-section} we present the solutions for the challenges we faced during the implementation of our algorithm.
Also, we show there the architecture of the system along with the system recovery algorithm. In Section~\ref{verification-section}, we discuss the usage of our algorithm to implement and verify the NVRAM CAS algorithm, along with the method of checking executions of the NVRAM CAS algorithm for serializability. In Section~\ref{future-section}, we discuss the directions of the future research. We conclude our work with Section~\ref{conclusion-section}.

\section{Model}
\label{model-section}

\subsection{System model}
Our system model is based on the model described in \cite{attiya2018nesting}.

\begin{figure}[H]
     \centering
     \begin{subfigure}[b]{0.45\linewidth}
          \centering
          \includegraphics[width=\linewidth]{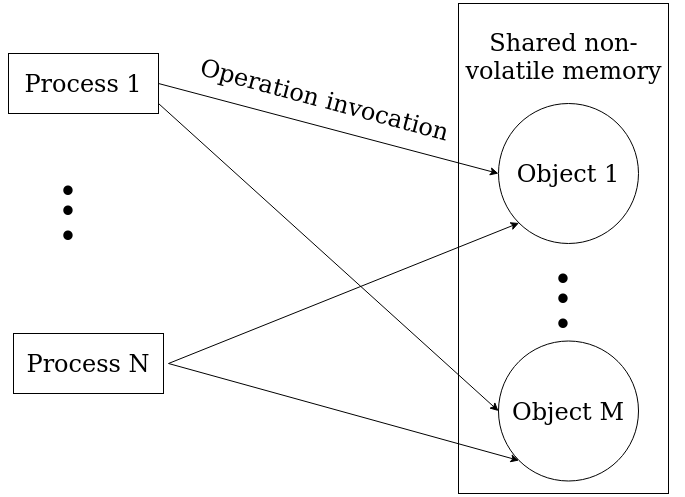}
          \captionof{figure}{Execution of operations on shared objects}
          \label{model-shared-pic}
     \end{subfigure}
     \hfill
     \begin{subfigure}[b]{0.45\linewidth}
         \centering
          \includegraphics[width=\linewidth]{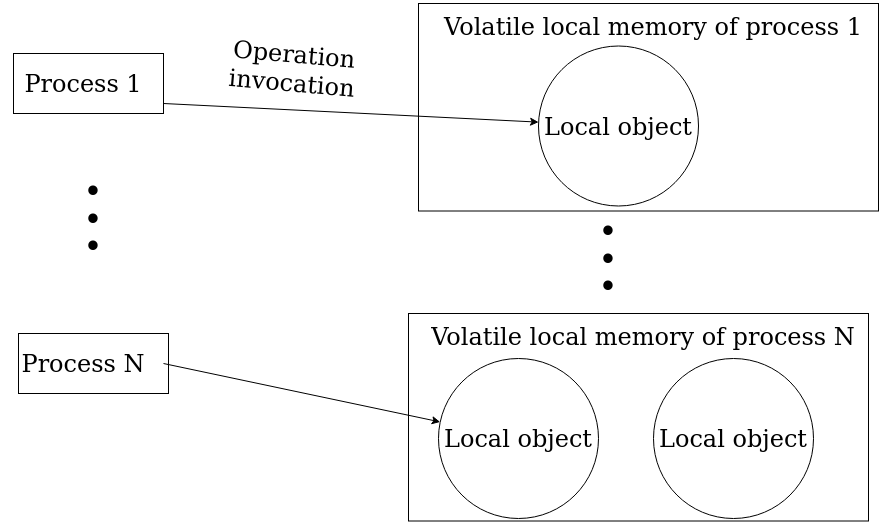}
          \captionof{figure}{Execution of operations on local objects}
           \label{model-local-pic}
     \end{subfigure}
        \caption{System execution model}
        \label{fig:three graphs}
\end{figure}

There are $N$ processes $\{p_i\}_{i = 1}^N$ executing operations concurrently. Also, there are $M$ objects $\{O_j\}_{j = 1}^M$ located in the shared non-volatile memory. Processes communicate with each other by executing operations on shared objects (see Fig.~\ref{model-shared-pic}), that can support \texttt{read}, \texttt{write} or \texttt{read-modify-write} \cite{herlihy1991wait} operations.

In our model, all shared memory is considered non-volatile, i.e., it does not lose its content even after a crash event. However, we assume the existence of a volatile memory in the system. Each object $LO$, located in the volatile memory, is considered local to some process $p$. In other words, only process $p$ can access object $LO$. Thus, besides being able to execute operations on shared objects, each process can access its local objects. Such objects support only \texttt{read} and \texttt{write} operations (see Fig.~\ref{model-local-pic}).

However, our model still does not reflect some properties of the real-world hardware: for example, it does not take into account the existence of the volatile NVRAM cache and the existence of shared volatile memory.

\subsection{Failure model}

There exist two general failure models:

\begin{itemize}
    \item \emph{Individual} crash-recovery model~\cite{attiya2018nesting}. In such a model, each process can face a crash event independently of all other processes. When a process faces a crash event, it stops working until it is restarted. All data, stored in the volatile memory of the failed process, is lost. However, all data persisted to the NVRAM is not lost and remains available to the failed process after its restart.

    \item \emph{System} crash-recovery model. In such a model, a crash event happens in the whole system instead of an individual process. The whole systems stops working until it is restarted. After the system restarts, the contents of all the volatile memory is lost. As in the previous model, the data, persisted in the NVRAM, is not lost and remains available to all processes after the system restarts.
\end{itemize}

Note, that the \emph{system} crash-recovery model is a special case of the \emph{individual} crash-recovery model, since a crash of the whole system can be represented as a set of $N$ simultaneous crash events of individual processes~--- one crash event per each process. 
Despite the fact that \emph{individual} crash-recovery model is a more general model, in this work we focus mainly on \emph{system} crash-recovery model. In real-world shared memory systems multiple computational units are placed in a single server and thus a failure of a single computational unit is impossible without a failure of the entire system. That is why, in our opinion, \emph{system} crash-recovery model describes more accurately the real-life crash event~--- for example, power loss.

\subsection{Operation execution}

We say that function \texttt{F} is being executed by process $p$ if execution of \texttt{F} has been started by $p$ but has not been finished yet. As described in \cite{attiya2018nesting}, we work with the nested invocation of functions: at any moment, multiple functions can be executed by any process. It happens when function \texttt{F} invokes function \texttt{G}. Thus, executed functions in each process form a nested sequence. In the above example execution of \texttt{G} is nested into the execution of \texttt{F}. 

To allow the recovery of the system, we provide each function \texttt{F} with a dual function \texttt{F.Recover}, which receives the same arguments as \texttt{F}. \texttt{F.Recover} is called after the system restart to perform system recovery and it should either finish the execution of \texttt{F} or roll back \texttt{F}.

To perform the system recovery, for each process $p$ we should call \texttt{F.Recover} for each function \texttt{F} being executed by $p$ at the crash moment. Moreover, recovery functions should be called in the certain order: if the execution of \texttt{G} is nested into the execution of \texttt{F}, \texttt{G.Recover} should be called before \texttt{F.Recover}. Thus, each process should perform the recovery in the LIFO (stack) order.

Also we should consider the possibility of \emph{repeated failures}~--- failures which happen during the recovery procedure. Consider the system failure after \texttt{F} was invoked. After the restart, we should call \texttt{F.Recover} to complete the recovery. Suppose another system failure happens before \texttt{F.Recover} is finished. After the second restart, we should again continue the recovery at executing \texttt{F.Recover}. It means that there is no difference between the system failure happening during the execution of \texttt{F} or during the execution of \texttt{F.Recover}: in both cases, we should call \texttt{F.Recover} to complete the recovery. Thus, \texttt{F.Recover} should be designed so that it can complete the operation (or roll it back) no matter whether the crash occurred when executing \texttt{F} or \texttt{F.Recover}.

\subsection{Correctness}
Multiple correctness conditions for NVRAM exist. Here, we outline three most important (from the strongest to the weakest):

\begin{enumerate}
    \item \emph{Nesting Safe Recoverable Linearizability} \cite{attiya2018nesting}. It requires each invoked function \texttt{F} to be completed even if a crash event occurs while executing \texttt{F}. Thus, under that correctness condition, \texttt{F.Recover} should finish the execution of \texttt{F} either by completing it successfully or by rolling it back.
     
    \item \emph{Durable Linearizability} \cite{izraelevitz2016linearizability}. It requires that each function \texttt{F}, execution of which has finished before a crash, should be completed. If a crash event occurs while executing function \texttt{F}, such function may be either completed or not.
    
    \item \emph{Buffered Durable Linearizability} \cite{izraelevitz2016linearizability}. It is a weaker form of \emph{Durable Linearizability}. Its difference is in that it allows function \texttt{F} not to be completed even if its execution finished before a crash. However, that correctness condition requires each object to provide \texttt{sync} operation~--- all functions, finished before a call to \texttt{sync} must be completed, even if a crash event occurs.
\end{enumerate}

In this work, we propose an algorithm that can be used to run NVRAM-destined programs under \emph{Nesting Safe Recoverable Linearizability}~--- the strongest correctness condition.

\section{Persistent Stack}
\label{stack-section}

\subsection{Program stack concept}

In order to execute programs for NVRAM, for each thread\footnote{When talking about practical aspects of concurrent programming, we use the word ``thread'' in the same context, as the word ``process'' in the theory of concurrent programming} $t$ we maintain an information about functions, which were executed by $t$ when a crash occurred. Also, to invoke recover functions in the correct order we maintain the order in which these functions were invoked.

We maintain that order by using the notion of program stack: each thread $t$ has its own NVRAM-located stack, and each function executed by $t$ corresponds to a single frame of the stack. When a function is invoked, the corresponding frame is added to the top of the stack. After the end of the execution, the frame is removed from the top. Therefore, when a crash occurs, the stack of thread $t$ contains frames, that correspond to functions that were executed by $t$ at the crash moment. Moreover, such frames are located in the correct order: if execution of \texttt{G} was nested into the execution of \texttt{F}, a frame of \texttt{G} is located closer to the top of the stack, than a frame of \texttt{F}.

\subsection{Issues of existing implementations}
\label{stack-issues}

The functionality of the program stack is already implemented by standard execution systems: for example, \texttt{x86} program stack. However, we cannot use them as-is, even if we transfer it from the  DRAM to the NVRAM. 

Here we remind the implementation of the function call via the \texttt{x86} stack~\cite{x86-call,x86-ret}. Suppose function \texttt{F} calls function \texttt{G} using \texttt{x86} command \texttt{CALL G}. To perform such an invocation, we should store a return address on the stack~--- the address of the instruction in function \texttt{F} that follows the instruction \texttt{CALL G}. After the execution of \texttt{G} is finished, we continue execution of \texttt{F} from that instruction. This is exactly how \texttt{x86} instruction \texttt{RET} works~--- it simply reads the return address from the stack and performs \texttt{JMP} \cite{x86-jmp} to that address, allowing it to the continue the execution from the desired point. 

Note that such a program stack implementation has a number of drawbacks, that makes it impossible for us to use such implementation as a persistent stack:

\begin{itemize}
    
    \item After the system restart due to the crash, the code segment may be relocated, i.e., have a different offset in the virtual address space. That will make us incapable of identifying which functions were executed at the crash moment~--- we simply won't be able to match return address from the stack with an address of some instruction after the code segment relocation.
    
    \item We cannot guarantee an atomicity of adding a new frame to the stack or removing a frame from the top of the stack~-- if a crash occurs during adding or removing stack frame, after the system restarts the stack might be in an inconsistent state.
\end{itemize}

Thus, instead of using existing program stack implementations, we present our persisted stack structure that overcomes the above drawbacks.

\subsection{Persistent stack structure}

Each thread has an access to its own persistent stack. For simplicity, in this section we assume that the persistent stack is allocated in the NVRAM as a continuous memory region of constant size. However, we explain how to make a stack of unbounded size in Appendix~\ref{inf-stack}.

\begin{figure}[H]
      \centering
      \includegraphics[width=0.75\linewidth]{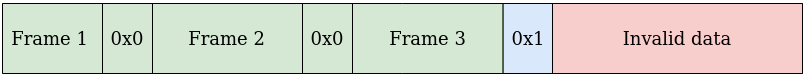}
      \captionof{figure}{Persistent stack structure}
      \label{stack-structure}
 \end{figure}

Persistent stack consists of consequent persistent stack frames~--- one frame per function that accesses NVRAM. \footnote{Each such function must have a recover version, as described above.} Each frame ends with a one-byte end marker: it is \texttt{0x1} (\emph{stack end marker}) if the frame is the last frame of the stack; otherwise, it is \texttt{0x0} (\emph{frame end marker}). Any data located after the \emph{stack end marker} is considered invalid~--- it should never be read or interpreted in any way (see Fig.~\ref{stack-structure}).

To finish the description of the data layout, each persistent stack frame consists of: 1)~a unique identifier of the invoked function that allows us to call the appropriate recover function during the system recovery; 2)~arguments of the function, serialized into a byte array---during the system recovery we pass them to the recover function; 3)~a one-byte end marker (either \texttt{0x0} or \texttt{0x1}).

\subsection{Update of the persistent stack}
 
The persistent stack should be updated: 1)~when the function is invoked~--- a new frame should be added to the top of the stack, 2)~when the function execution is finished~--- the top frame of the stack should be removed.
 
\subsubsection{Adding the new frame to the top of the stack.}
\label{frame-add}

\begin{figure}[H]
     \centering
     \begin{subfigure}[b]{0.75\linewidth}
          \centering
          \includegraphics[width=\linewidth]{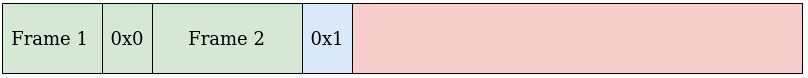}
          \captionof{figure}{Persistent stack before the function invocation}
          \label{stack-before-add}
     \end{subfigure}
     \hfill
     \begin{subfigure}[b]{0.75\linewidth}
         \centering
          \includegraphics[width=\linewidth]{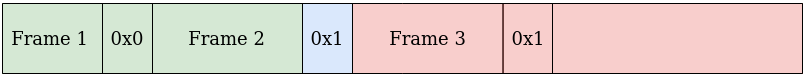}
          \captionof{figure}{Persistent stack after writing the new frame after the stack end marker}
          \label{stack-add-in-progress}
     \end{subfigure}
     \begin{subfigure}[b]{0.75\linewidth}
         \centering
         \includegraphics[width=\linewidth]{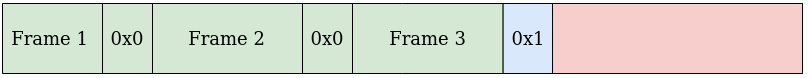}
         \captionof{figure}{Persistent stack after adding the new frame to the top of the stack}
         \label{stack-after-add}
     \end{subfigure}
        \caption{Adding new frame to the top of the stack}
        \label{adding-frame}
\end{figure}

Suppose the stack at the beginning of the operation has two frames in it (see Fig.~\ref{stack-before-add}):

To add a new frame to the top of the stack, we perform the following actions:

\begin{enumerate}
    \item After the \emph{stack end marker}, we write a new frame with the \emph{stack end marker} set. Note that the new frame (frame 3) is located after the \emph{stack end marker} of the previous frame (frame 2). Therefore, the new frame is not considered as a stack frame, while the previous frame (frame 2) is still the last stack frame (see Fig.~\ref{stack-add-in-progress}). 
    
     \item Change the end marker of the current last stack frame (frame 2) from \texttt{0x1} to \texttt{0x0}. Thus, the last stack frame (frame 2) becomes the penultimate stack frame and the new frame (frame 3) becomes the last stack frame (see Fig.~\ref{stack-after-add}). We name that one-byte end marker changing operation as \emph{moving the stack end forward}.
\end{enumerate}

\subsubsection{Removing the top frame from the stack.}
\label{frame-remove}

\begin{figure}[H]
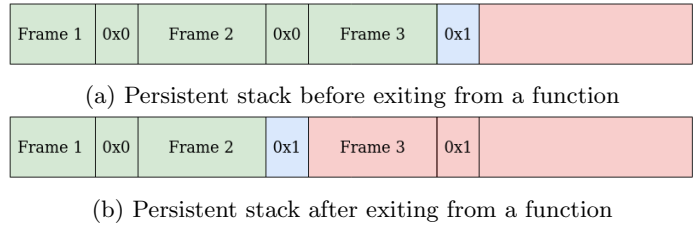

     \centering
     \begin{subfigure}[b]{0.75\linewidth}
          \centering
          \includegraphics[width=\linewidth]{stack-after-add.png}
          \captionof{figure}{Persistent stack before exiting from a function}
          \label{stack-before-exit}     \end{subfigure}
     \hfill
     \begin{subfigure}[b]{0.75\linewidth}
         \centering
          \includegraphics[width=\linewidth]{stack-add-in-progress.png}
          \captionof{figure}{Persistent stack after exiting from a function}
          \label{stack-after-exit}
          \end{subfigure}
        \caption{Removing the top frame from the stack}
        \label{removing-frame}
\end{figure}

Suppose the stack at the beginning of the operation has three frames in it (see Fig.~\ref{stack-before-exit}):

To remove the top frame from the stack, we simply change the end marker of the penultimate stack frame (frame 2) from \texttt{0x0} to \texttt{0x1}, thus making the penultimate stack frame the last stack frame (see Fig.~\ref{stack-after-exit}). We name that one-byte end marker changing operation as \emph{moving the stack end backward}. Note, that frame 3 becomes the part of the invalid data and, therefore, it will not be considered as a stack frame anymore.
 
\subsubsection{Dummy frame.}

Note that both frame removal and frame addition procedures assume the existence of at least one frame in the stack, besides the one that is being removed or added. Particularly, this assumption implies that the bottom stack frame cannot be removed from the stack. We can simply satisfy that assumption by introducing a dummy frame~--- the stack frame, located at the bottom of the stack (i.e. the first frame, added to the stack). That frame is added to the stack at the initialization of the stack and is never removed. By that, we ensure that there is always at least one frame, thus making it possible for us to use the stack update procedures, described above.
 
\subsubsection{Flushing long frames.}
\begin{figure}[H]
    \centering
    \includegraphics[width=0.75\linewidth]{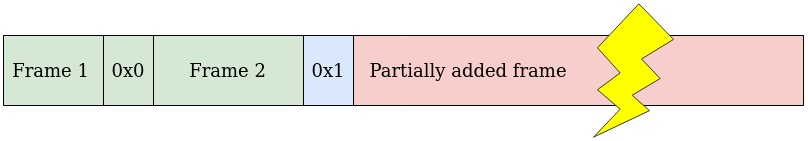}
    \captionof{figure}{Persistent stack with partially flushed frame}
    \label{stack-broken-frame}
 \end{figure}

Note that sometimes a new stack frame does not fit into a single cache line~--- for example, that can happen when some function receives arguments list with length greater than the cache line size. In such case, we will not be able to add such frame to the stack atomically (since only single cache line can be persisted atomically). Therefore, we can face a crash event that will force us to write the new frame partially (see Fig.~\ref{stack-broken-frame}).

In our algorithm, we, at first, add a new frame to the stack, and only after the new frame has been written successfully we \emph{move the stack end forward}. Thus, even if the crash event happens, the stack will remain consistent: partially written frame will be located after the \emph{stack end marker} and will not be considered as a stack frame. Therefore, this scenario does not brake \emph{Nesting-safe Recoverable Linearizability}, since the last function invocation was not linearized before the crash event. We can simply think that the crash happened before the function invocation and the function was never invoked.

\subsubsection{The atomicity of the stack update.}

We can say that a function invocation linearizes only when we \emph{move the stack end forward}. This requires only the flushing of a single byte to the NVRAM. Since a single byte always resides in a single cache line, this flush always happens atomically.

The same observation can be made for \emph{moving the stack end backward}: an execution of the function is finished when we change the end marker of the penultimate stack frame from \texttt{0x0} to \texttt{0x1}. As was described above, such action happens atomically.

\subsubsection{Persistent stack and the NVRAM}

The procedure of adding and removing a stack frame requires only the ability to flush a single byte atomically and not the entire cache line~--- this makes us capable of implementing the stack maintenance algorithm on a hardware that does not support atomic flushing of an entire cache line. Thus, the algorithm described above, can be easily emulated without having access to an expensive NVRAM hardware, using almost any existing persistent hardware such as HDD, SSD, etc. 

\begin{figure}[H]
     \centering
     \begin{subfigure}[b]{0.57\linewidth}
          \centering
          \includegraphics[width=0.75\linewidth]{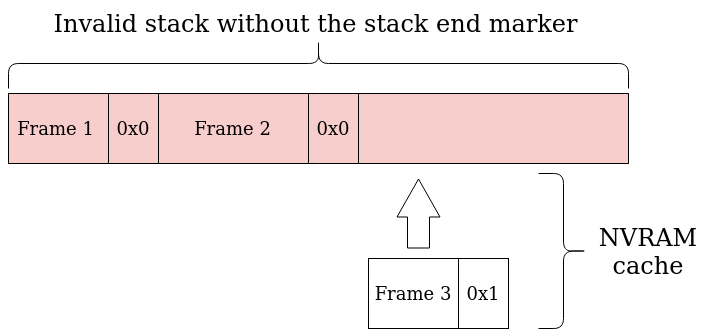}
        \captionof{figure}{New stack frame has been lost due to volatility of the NVRAM cache}
        \label{stack-cache-frame}
     \end{subfigure}
     \hfill
     \begin{subfigure}[b]{0.39\linewidth}
         \centering
          \includegraphics[width=\linewidth]{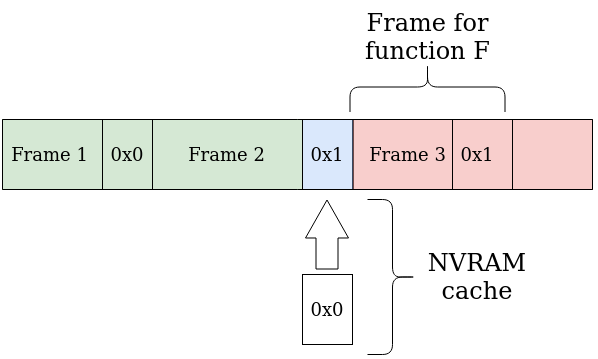}
          \captionof{figure}{End marker has been lost due to volatility of the NVRAM cache}
        \label{stack-cache-marker}
     \end{subfigure}
        \caption{Results of violating flushing invariants}
        \label{fig:three graphs}
\end{figure}

For the above reasoning to remain correct, we should maintain two following invariants:

\begin{enumerate}
    \item We should flush the new stack frame before \emph{moving the stack end forward}.
    
    Suppose we violate that rule. Consider a crash event that happens at some time after the \emph{moving the stack end forward}. Suppose also, that new stack frame (frame 3) has been written to the volatile NVRAM cache and was lost during the crash. After the system restart we will not be able to call the recover function for frame 3, because we have lost that frame (see Fig.~\ref{stack-cache-frame}).
    
    \item When changing the end marker of some frame (either from \texttt{0x0} to \texttt{0x1} or vice versa) we should immediately flush it before staring the execution of the invoked function or continuing the execution of the caller function. 
    
    Suppose we violate that rule. Consider a crash event, happening while executing function \texttt{F}, corresponding to frame 3. Also consider that the \emph{frame end marker}, written to frame 2, has been written to the volatile NVRAM cache and thus has been lost (see Fig.~\ref{stack-cache-marker}). After the system restart, we do not consider frame 3 as a stack frame, and, thus, we do not even invoke \texttt{F.Recover}.
\end{enumerate}

\section{System implementation}
\label{system-section}

\subsection{Pointers to the memory in NVRAM}
\label{nvram-pointers}

When working with pointers to the NVRAM we face the problems similar to those we faced when working with function addresses (Section \ref{stack-issues}). Suppose we have acquired pointer \texttt{ptr} pointing to the NVRAM. We store \texttt{ptr} in the NVRAM (for example, in some persistent stack frame, as an argument of some function \texttt{F}). After that, we face a crash event. And when we restart the system, the mapping of the NVRAM into the virtual address space can change, thus, making pointer \texttt{ptr} invalid, since it does not point to the NVRAM anymore.

The same problem happens when we emulate NVRAM using HDDs, mapped to the virtual address space using \texttt{mmap} syscall \cite{mmap}: on each system restart, HDD is mapped to a different location in the virtual address space.

This problem has a very simple solution: instead of using direct pointers to the NVRAM, we shall use offsets from the beginning of the NVRAM mapping into the virtual address space. Suppose the mapping of the NVRAM begins at address \texttt{MAP\_ADDR}. Then, instead of storing \texttt{ptr} we store \texttt{ptr - MAP\_ADDR}~--- an offset of the desired memory location. Note that such an offset does not depend on an exact location of the mapping, thus making it safe for us to store it in the NVRAM and use after the system restart.

\subsection{Handling return values}

Traditionally, on \texttt{x86} architecture, functions return value using the volatile memory~--- either in \texttt{x86} register \texttt{EAX}, if the return value is an integer, or in \texttt{FPU} register \texttt{ST0}, if the return value is floating-point. For example, \texttt{cdecl}, one of the most popular \texttt{x86} calling conventions \cite{calling-conventions}, implies the above rules for return value.

However, in our case we cannot use volatile processor registers to store return value. Consider a crash event occurring after the callee function \texttt{G} has saved the return value to the \texttt{EAX} and finished its execution by \emph{moving the stack end backward}. At that time, the caller function \texttt{F} has not persisted the return value from \texttt{EAX} to the NVRAM. After the system restart, we will not invoke \texttt{G.Recover}, but start from \texttt{F.Recover} instead. However, we cannot execute \texttt{F.Recover} properly, because we have lost the result of \texttt{G}.

That is why functions should store their results directly in the NVRAM. We could come up with two approaches where to store them:

\begin{enumerate}
    \item on the persistent stack. For example, we can use an especially-allocated place in a persistent stack frame for that purpose.
    
    \item in the NVRAM heap. In such a case, the caller can preallocate a memory location for the answer before invoking the callee, and pass the pointer to that memory location in callee's arguments (note, that as was mentioned in Section~\ref{nvram-pointers}, we should use offsets instead of pointers to the NVRAM). After that, callee can store its answer in that memory location.
\end{enumerate}

In both cases, the callee should flush the answer to the NVRAM before \emph{moving the stack end backward}. Our implementation supports returning of small values (up to 8 bytes) on the persistent stack, while big values are returned in the NVRAM heap.

\subsection{Architecture of the system}

The system consists of a single main thread and \texttt{N} worker threads. 

Main thread can run in either a standard mode or a recovery mode.

When running in the standard mode, the main thread performs the following steps.

\begin{enumerate}
    \item Initialize the NVRAM heap. This may include the initialization of the memory allocator, the mapping of the NVRAM to the virtual address space and the consequent initialization of the variable \texttt{MAP\_ADDR}, mentioned in Section~\ref{nvram-pointers}
    
    \item Initialize \texttt{N} new persistent stacks.
    
    \item Start \texttt{N} worker threads, giving each worker thread pointer to the beginning of its persistent stack.
    
    \item Receive task that should be executed by the system and add them to the producer-consumer queue.
\end{enumerate}

When running in the standard mode, worker threads receive tasks from the producer-consumer queue and execute them.

In case of a crash, the main thread starts in the recovery mode and performs the following steps:

\begin{enumerate}
    \item Initialize the NVRAM heap.
    
    \item Start \texttt{N} recovery threads, giving each recovery thread the pointer to a persistent stack of some worker thread.
    
    \item Wait for all recovery threads to finish.
    
    \item Restart the system in normal mode.
\end{enumerate}

Each recovery thread executes the following algorithm:

\begin{enumerate}
    \item Traverse its persistent stack from the top to the bottom.
    
    \item Execute the corresponding recover operation for each stack frame.
    
    \item After the recovering of an operation on the top of the stack is finished, pop the top frame.
    
    \item After all the frames (except for the dummy one) are removed from the stack, finish the execution.
\end{enumerate}

System recovery happens in parallel, which allows for a faster recovery than an ordinary single-threaded recovery.

We note that our algorithm deals well with \emph{repeated failures}. If such a failure happens during the recovery, the new recovery continues not from the beginning, but from where the previous recovery was interrupted. More formally, consider a frame, corresponding to a function \texttt{F}. If during the recovery we have completed execution of \texttt{F.Recover} and removed that frame from the stack, even after the \emph{repeated failure} we will not run a recover function for that frame once more. Thus, we achieve the progress even in the presence of \emph{repeated failures}.

\section{Verification}
\label{verification-section}

The described algorithm of the persistent stack can be used to implement and verify \texttt{CAS} algorithm for NVRAM, described in \cite{attiya2018nesting}. That paper assumes the absence of the volatile NVRAM cache, i.e., all writes are performed right into the memory. To emulate this, we should flush each written cache line to the NVRAM immediately after the corresponding write. Also, we should implement the algorithm so that each written value never crosses the border of a cache line to allow atomic flush of each written value.

Multiple correctness conditions for concurrent algorithms exist: the most popular are linearizability \cite{herlihy1990linearizability}, sequential consistency and serializability \cite{papadimitriou1979serializability}. We want to perform the verification against some of these correctness conditions.

From now on we take \texttt{CAS} algorithm for the NVRAM as the running example. Consider the following execution. Multiple threads run a set of \texttt{CAS} operations on a single register \texttt{Reg}: $\{CAS(Reg, old_i, new_i)\}_{i = 1}^N$, and the initial value of \texttt{Reg} is \texttt{init}. And for each operation we know whether it was finished successfully or not.

We present an algorithm, that can be used to check such an execution for serializability in a polynomial time.

\subsection{Serializability}

To verify the execution for serializability in polynomial time, we build a graph $\langle V, E \rangle, G = \{old_i\}_{i = 1}^N \cup \{new_i\}_{i = 1}^N \cup \{init\}$ and construct the set of edges $E$ the following way: $a \rightarrow b \in E$ if and only if there exists a successful $CAS(Reg, a, b)$ in the execution. Also, we read the final value of the register. We can read it after all the \texttt{CAS} operations are finished. 

Since each edge of $G$ corresponds to a successful \texttt{CAS}, each successful CAS was executed exactly once, and each successful $CAS(Reg, a, b)$ changed value of $Reg$ from $a$ to $b$, to verify the execution for serializability we should find some Eulerian circuit that starts in the initial value of the register and ends in the final value of the register~--- such a circuit corresponds to the sequential execution. Thus, the execution is serializable if and only if such a circuit can be found\footnote{Please, note that we can simply serialize unsuccessful operation at the times when the register holds a value different from $old_i$.}.

\subsection{Running examples}

We have implemented the algorithm, described above, using HDD-based memory-mapped files \cite{mmap} to emulate the NVRAM. We used UNIX utility \texttt{kill} \cite{kill} to interrupt the system at random moments by that emulating system crashes.

We have generated random executions of the algorithm in the following way:

\begin{enumerate}
    \item Generate an initial integer value of the register;
    
    \item Generate $\{new_i\}_{i = 1}^N$ and $\{old_i\}_{i = 1}^N$ as integer values, uniformly sampled from some range: either wide range $[-10^5, 10^5]$), or narrow range ($[-10, 10]$);
    
    \item Start the system in the normal mode, add descriptors of $\{CAS(Reg, old_i, new_i)\}_{i = 1}^N$ operations to the producer-consumer queue in the random order;
    
    \item Run $4$ working threads that execute these \texttt{CAS} operations;
    
    \item At random moment, emulate system failure using the \texttt{kill} utility;
    
    \item Restart the system in the recovery mode waiting for all \texttt{CAS} operations, that were executing at the crash moment, to complete;
    
    \item Restart the system in the normal mode, add all remaining descriptors to the queue;
    
    \item Run steps 4-7 until all operations are completed;
    
    \item Get answers of all \texttt{CAS} operations, get the final value of the register, and, finally, verify the execution for serializability.
\end{enumerate}

We have verified a lot of random executions along with emulated system failures at random moments. All executions of the CAS algorithm presented in \cite{attiya2018nesting} were found to be serializable. We also verified the executions of incorrect CAS algorithm with especially-added bugs: we have removed the matrix \texttt{R} from the \texttt{CAS} algorithm. The executions of such a wrong implementation were reported to be non-serializable.

The implementation is publicly available at \url{https://github.com/KokorinIlya/NVRAM\_runner}.

\section{Future work}
\label{future-section}

We find three interesting directions for the future work: 1)~implement and test other NVRAM algorithms; 2)~find the polynomial algorithm that verifies executions of \texttt{CAS} algorithm for linearizability and sequential consistency, or prove that the problem of such a verification is NP-complete; 3)~develop a plugin for one of the modern C++ compilers that can be used to reduce the boilerplate code: e.g., automatically create a new stack frame on each function call, remove the top frame when a function execution finishes, and etc.

\section{Conclusion}
\label{conclusion-section}

In this paper we presented an algorithm that can be used to run NVRAM programs. The described algorithm takes into consideration different aspects of real-world NVRAM systems. Moreover, the algorithm can be used to run NVRAM-destined programs on commodity persistent hardware, which can be useful for implementing and testing novel NVRAM algorithms without having an access to an expensive NVRAM hardware. The algorithm was successfully used to implement and verify the CAS algorithm for NVRAM.

%
%
%
%
\bibliographystyle{splncs04}
\bibliography{references}

\appendix

\section{Stack of unbounded size}
\label{inf-stack}

\subsection{Motivation}

Sometimes we do not know the size of the program call stack in advance, so we have to make it to be resizable. Consider the following example.
We want to consequently update a lot of separate data items $\{a_i\}_{i = 1}^N$ in the transactional behaviour: if we update only a part of the requested items and face a crash event---after the system restart all modifications should be rolled back. In short, we want to make a transactional for-loop.

Luckily, we can represent a for-loop using the recursive function \texttt{F(i)}. This function saves an old value of $a_i$, performs an update of $a_i$ and calls \texttt{F(i + 1)}. A function \texttt{F.Recover(i)} rolls back an update of $a_i$.


Therefore, if we face a crash event during the update procedure, we can restore the system into the consequent state.

However, using such an implementation we face a problem with the constant size of the persistent stack: if the number of updated items $N$ is big, we cannot place frames of all update functions $\{\texttt{F(i)}\}_{i = 1}^N$ on the stack.

That is why we need to implement the persistent stack of unbounded size. We achieve it in two different ways: using either the resizable array or linked list.

\subsection{Dynamically resizable array}

\begin{figure}[H]
    \centering
    \includegraphics[width=\linewidth]{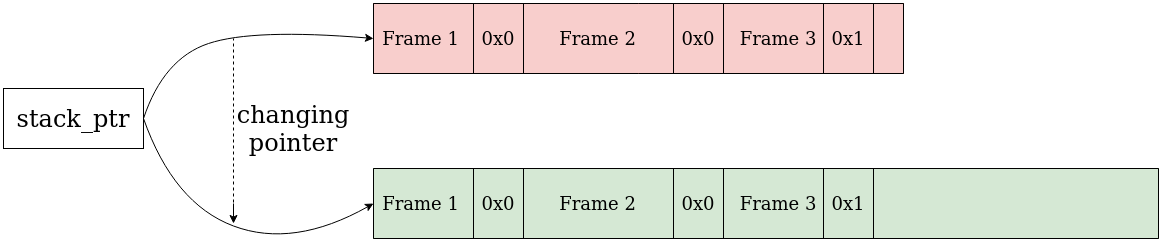}
    \captionof{figure}{Persistent stack using dynamically resizable array}
    \label{stack-array}
\end{figure}

The first approach to this problem is based on the idea of a dynamically resizable array. Along with the continuous block we allocate a single pointer in the NVRAM that points to the block storing the stack data (see Fig.~\ref{stack-array}). 

We refer to the length of that continuous block in bytes as the \emph{capacity} of the stack, while referring to the length of all frames in bytes as \emph{size} of the stack.

Sometimes we need to change the capacity of the stack. This can happen in two situations: either we want to add a new frame to the stack and it does not fit into the current block, or we think that capacity is too big (e.g. \texttt{capacity > 4 $\cdot$ size}) and we want to shrink the stack.

To do so, we execute the following steps: 1)~allocate a new memory block in the NVRAM with the desired capacity; 2)~copy the stack data from the old block into the new block; 3)~flush the copied data; 4)~change the pointer to point onto the new block; 4)~flush the pointer to the NVRAM.

After that, we deallocate the old block and continue with the stack as usual.

The main disadvantage of this method is that on the resizing it requires $O(\texttt{size of the stack})$ time to copy the contents from the old block to the new block.

\subsection{Linked list of blocks}

The second approach uses a linked list of blocks (see Fig.~\ref{stack-list-before-remove}). Each block contains a pointer to the next block in the list, pointer to the first block is stored in some well-known location.

We introduce two types of frames. The ones of the first type are the \emph{ordinary frames} as introduced before: they correspond to some function, being executed by current thread, and contain an identifier of that function and its arguments. The ones of the second type are \emph{pointer frames}: they contain a pointer to the next block in the linked list \footnote{As was said in section \ref{nvram-pointers}, we should use offsets instead of pointers to the NVRAM}. To distinguish one frame from another we may augment each frame with one-byte preamble: ordinary frames begin with \texttt{0xA}, while pointer frames begin with \texttt{0xB}.

When we encounter the pointer frame at the current block, we seek the next frame at the block pointed at by the pointer from the pointer frame. All the data from the current block after the pointer frame should be considered invalid.

\subsubsection{Adding a new frame to the top of the stack}

If the new frame fits into the current block, we add the new frame as we did it in Section~\ref{frame-add}. Otherwise, we execute the following steps: 1)~allocate a new block that accommodates the new frame; 2)~add a pointer frame, pointing to the new block, to the top of the current block.

\subsubsection{Removing a frame from the top of the stack}

\begin{figure}[H]
     \centering
     \begin{subfigure}[b]{0.75\linewidth}
          \centering
          \includegraphics[width=\linewidth]{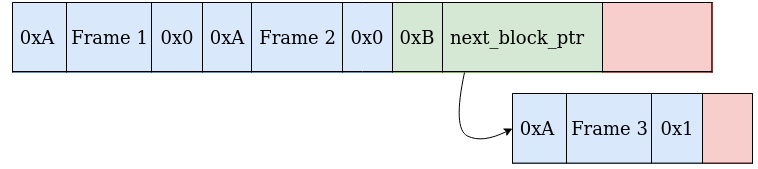}
          \captionof{figure}{Persistent stack as a list of blocks}
          \label{stack-list-before-remove}     \end{subfigure}
     \hfill
     \begin{subfigure}[b]{0.75\linewidth}
         \centering
          \includegraphics[width=\linewidth]{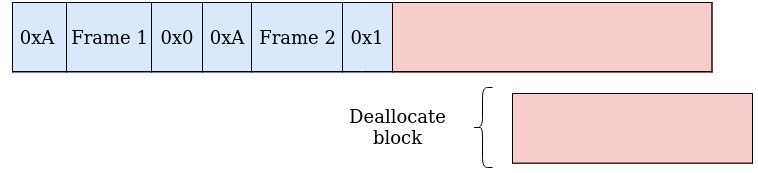}
          \captionof{figure}{Persistent stack after removing the top frame}
          \label{stack-list-after-remove}
          \end{subfigure}
        \caption{Removing the top frame from the stack}
        \label{stack-list-remove}
\end{figure}

If the top frame of the stack is not the only frame of the current block, we remove it as in Section~\ref{frame-remove}. Otherwise, we execute the following steps: 1)~find the previous block;  
2)~in the previous block delete the last stack frame, which is the pointer frame, pointing at the current block; 3)~after that, we deallocate the unnecessary block (see Fig.~\ref{stack-list-remove}). 

In order to be able to quickly fetch the previous block of the current block we can use any of these two approaches:

\begin{itemize}
    \item Use a double linked list of blocks in which each block contains a pointer to its previous block;
    
    \item Load addresses of all blocks in some container with random access (e.g., dynamic array).
\end{itemize}

\end{document}